\documentclass[ste,twoside]{stefano}
\usepackage{cite}
\usepackage{amssymb}
\usepackage{graphics}
\usepackage{rotating}

\usepackage{amsmath}

\def\makeheadbox{{%
\hbox to0pt{\vbox{\baselineskip=10dd\hrule\hbox
to\hsize{\vrule\kern3pt\vbox{\kern3pt \hbox{ {\sc quaternionic
eigenvalue problem} } \hbox{ {\sf Journal of Mathematical Physics {\bf 43},
5815-5829 (2002)}
\hspace*{5.8cm} $\boldsymbol{\Sigma \delta \Lambda}$ }
\kern3pt}\hfil\kern3pt\vrule}\hrule}%
\hss}}}

\def\0{\mbox{\tiny $0$}}
\def\1{\mbox{\tiny $1$}}
\def\2{\mbox{\tiny $2$}}
\def\3{\mbox{\tiny $3$}}
\def\4{\mbox{\tiny $4$}}
\def\5{\mbox{\tiny $5$}}
\def\6{\mbox{\tiny $6$}}
\def\7{\mbox{\tiny $7$}}
\def\8{\mbox{\tiny $8$}}
\def\9{\mbox{\tiny $9$}}
\def\X{\mbox{\tiny $\mathbb{X}$}}
\def\Re{\mbox{\tiny $\mathbb{R}$}}
\def\Co{\mbox{\tiny $\mathbb{C}$}}
\def\Ha{\mbox{\tiny $\mathbb{H}$}}
\def\x{\mbox{\tiny $x$}}

\begin{document}

\title{QUATERNIONIC EIGENVALUE PROBLEM}

\author{
Stefano De Leo\inst{1}
%\thanks{Partially supported by ...}
\and
Giuseppe Scolarici\inst{2}
\and
Luigi Solombrino\inst{2}
%\thanks{Supported by ...}
}

\institute{
Department of Applied Mathematics, University of Campinas\\
PO Box 6065, SP 13083-970, Campinas, Brazil\\
{\em deleo@ime.unicamp.br}
\and
Department of Physics, University of Lecce and INFN, Sezione di Lecce\\
PO Box 193, 73100, Lecce, Italy\\
{\em scolarici@le.infn.it}\\
{\em solombrino@le.infn.it}
}

%%%%%%%%%%%%%%%%%%%%%%%%%%%%%%%%%%%%%%%%%%%%%%%%%%%%%%%%%%%%%%%%%%%%%%%%%%%
%%%%%%%%%%%% DATE ABSTRACT PACS % %%%%%%%%%%%%%%%%%%%%%%%%%%%%%%%%%%%%%%%%%

\date{{\em July, 2002}}
% Warning: Where is the date?

\abstract{We discuss the (right) eigenvalue equation for
$\mathbb{H}$, $\mathbb{C}$ and $\mathbb{R}$ linear quaternionic
operators. The possibility to introduce an isomorphism between
these operators and real/complex matrices allows to translate the
quaternionic problem into an {\em equivalent} real or complex
counterpart. Interesting applications are found in solving
differential equations within quaternionic formulations
of quantum mechanics.}

%%%%%%%%%%%%%%%%%%%%%%%%%%%%%%%%%%%%%%%%%%%%%%%%%%%%%%%%%%%%%%%%%%%%%%%
%%%%%%%%%%%%%%%%%%%%%%%%%%%%%%%%%%%%%%%%%%%%%%%%%%%%%%%%%%%%%%%%%%%%%%%

% Warning: No PACS code given

% 02.10.Tq Associative rings and algebras
% 02.30.Hq Ordinary differential equations
% 02.30.Jr Partial differential equations
% 02.30.Tb Operator theory
% 03.65.-w Quantum mechanics

%\offprints{~Stefano De Leo.}

%\titlerunning{Quaternionic eigenvalue problem}

\maketitle

%%%%%%%%%%%%%%%%%%%%%%%%%%%%%%%%%%%%%%%%%%%%%%%%%%%%%%%%%%%%%%%%%%%%%%%%%%

\section*{I. INTRODUCTION}
\label{sec1}

The full understanding of the subtleties of the quaternionic
eigenvalue problem still represents an intriguing challenge for
mathematicians and physicists. The recent study of the eigenvalue
problem for complex linear quaternionic operators~\cite{Del00}
played a fundamental role in solving quaternionic differential
equations~\cite{DeDu}. In the last few years, interesting
applications of quaternionic analysis and linear algebra were
investigated in quantum mechanics~\cite{ADL}. In particular, the
solution of the Schr\"odinger equation in presence of quaternionic
perturbations was explicitly given for constant potentials and
deviations from standard (complex) quantum mechanics
discussed~\cite{DeDuNi}. In this paper, we aim to complete the
study begun in ref.~\cite{Del00} where preliminary steps in
solving the eigenvalue problem for complex linear quaternionic
operators were traced. In order to extend to the
$\mathbb{R}$-linear case the results obtained for the  $\mathbb{H}$ and
$\mathbb{C}$-linear quaternionic matrices, we have to introduce a
system of {\em coupled} equations which represents the {\em new}
eigenvalue problem for $\mathbb{R}$ linear quaternionic operators.
It is important to observe that no attempt to develop a complete
theory of the quaternionic eigenvalue problem has been made here,
this exceeds the scope of our paper. A  satisfactory discussion of
the eigenvalue problem for quaternionic operators is at present
far from being given. We could have directly investigated the
eigenvalue equation in the quaternionic space, but we have
preferred a more practical approach and chosen to handle the
problem by finding a more familiar real or complex space
isomorphic to the quaternionic one. We shall show that the
isomorphism between $\mathbb{H}$, $\mathbb{C}$, and
$\mathbb{R}$-linear quaternionic operators and real/complex matrices
immediately allows to translate the quaternionic (right)
eigenvalue problem in a corresponding real or complex counterpart.
The study of the new translated problem gives important
information about the quaternionic solution. The results obtained
are very useful in solving polynomial and differential equations.
This could represents a fundamental step in understanding the
potentiality of using quaternions in formulating quantum mechanics
(by investigating {\em quaternionic} deviations from the standard
theory~\cite{ADL,DeDuNi}) and gauge theory (by suggesting {\em
new} unification groups~\cite{QF}).

Throughout the paper we shall denote by $\mathbb{R}$, $\mathbb{C}$
and $\mathbb{H}$ the sets of real, complex and quaternionic
numbers, $\mathbb{R} \subset \mathbb{C} \subset \mathbb{H}$, and
by $V[n,\mathbb{X}]$ and $M[n,\mathbb{X}]$, respectively,  the
$n$-tuples and the $n \times n$ matrices over $\mathbb{X}$. Linear
quaternionic operators will be distinguished by their linearity
from the right. In what follows, the notation $\mathcal{O}_{\X}$
stands for quaternionic operators linear (from the right) over the
field $\mathbb{X}$.

\section*{II. QUATERNIONIC ALGEBRA AND LINEAR OPERATORS}
\label{sec2}

We now introduce the quaternionic algebra and  some useful
properties of $\mathbb{H}$, $\mathbb{C}$ and $\mathbb{R}$-linear
operators. The (real) quaternionic skew-field $\mathbb{H}$ is an
associative (division) algebra of rank 4 over $\mathbb{R}$,
\begin{equation}
q=q_{0}+ i \, q_{1} + j \, q_{2}+ k \, q_{3}~,~~~
q_{0,1,2,3} \in \mathbb{R}~,
\end{equation}
where
\begin{equation}
i^2=j^2=k^2=ijk=-1~,
\end{equation}
endowed with an involutory anti-automorphism ({\em conjugation})
\[
q \, \rightarrow \, \overline{q}=q_{0} - i \, q_{1} - j \, q_{2} - k \, q_{3}~.
\]
Due to the non-commutative nature of quaternions, we must
distinguish between the left and right action of the quaternionic
imaginary units $i$, $j$ and $k$. To do it, we introduce the
operators
\[
L_{\mu}=(1,L_{i},L_{j},L_{k})~~~\mbox{and}~~~
R_{\mu}=(1,R_{i},R_{j},R_{k})~,~~~\mbox{\small $\mu=0,1,2,3$}~,
\]
which act on quaternionic vectors $\psi \in V[n,\mathbb{H}]$ in
the following way
\begin{equation*}
L_\mu \psi =
h_{\mu} \psi~~~\mbox{and}~~~
R_{\mu} \psi
=\psi \, h_{\mu}~,~~~h_{\mu}=(1,i,j,k)~.
\end{equation*}
These operators satisfy
\begin{equation}
L_{i}^{2}=L_{j}^{2}=L_{k}^{2}=L_{i}L_{j}L_{k}=R_{i}^{2}=
R_{j}^{2}=R_{k}^{2}=R_{k}R_{j}R_{i}= -1~,
\end{equation}
and
\begin{equation}
\left[ \, L_{\mu} \, , \, R_{\nu} \, \right] = 0~,~~~
\mbox{\small $\mu,\nu=0,1,2,3$}~.
\end{equation}
Note that $\mathbb{H}$-linear quaternionic operators acting on a
finite $n$-dimensional quaternionic vector space,
\[
\mathcal{O}_{\Ha} \left( \psi_1 q_1 + \psi_2 q_2 \right) = \left(
O_{\Ha} \psi_1 \right) q_1 + \left( O_{\Ha} \psi_2 \right)
q_2~,~~~q_{1,2} \in \mathbb{H}~,~~~ \psi_{1,2} \in
V[n,\mathbb{H}]~,
\]
are in one to one correspondence with $n\times n$ quaternionic
matrices:
\begin{equation}
\mathcal{O}_{\Ha} ~\leftrightarrow~ M_{\Ha} \in M[n,\mathbb{H}]~.
\end{equation}
Consequently, $\mathbb{R}$ and $\mathbb{C}$-linear quaternionic
operators~\cite{Hor84}
\begin{eqnarray*}
\mathcal{O}_{\Re} \left( \psi_1 r_1 + \psi_2 r_2 \right) &=&
\left( \mathcal{O}_{\Re} \psi_1 \right) r_1 + \left(
\mathcal{O}_{\Re} \psi_2 \right) r_2~,~~~r_{1,2} \in \mathbb{R}~,
~~~\psi_{1,2} \in V[n,\mathbb{H}]~,\\
 \mathcal{O}_{\Co} \left( \psi_1 c_1 + \psi_2 c_2 \right) &=&
\left( \mathcal{O}_{\Co} \psi_1 \right) c_1 + \left(
\mathcal{O}_{\Co} \psi_2 \right) c_2~,~~~ \, c_{1,2} \in
\mathbb{C}~, ~~~\psi_{1,2} \in V[n,\mathbb{H}]~,\,
\end{eqnarray*}
can be represented by
 $n\times n$ quaternionic matrices $M_{\Ha}$
and right acting operators $R_{\mu}$ as follows
\begin{equation}
\mathcal{O}_{\Re} ~\leftrightarrow~M_{\Re} =  \sum_{\mu=0}^{3}
M_{\mu,\Ha} \, R_{\mu}~~~
\mbox{and}~~~
\mathcal{O}_{\Co} ~ \leftrightarrow ~ M_{\Co}  =  \sum_{s=0}^{1}
M_{s,\Ha} \,  R_{s}~.
\end{equation}
Thus, $\mathbb{R}$-linear  quaternionic operators consist of right
multiplication by quaternionic numbers ({\small $\mu = 0,1,2,3$})
whereas $\mathbb{C}$-linear  quaternionic operators are restricted
to right multiplication by complex numbers ({\small $s = 0,1$}).

\section*{III. THE EIGENVALUE PROBLEM}
\label{sec3}

In this section, we briefly discuss the left and right eigenvalue
equation for $\mathbb{H}$, $\mathbb{C}$ and $\mathbb{R}$-linear
quaternionic operators. As explicitly shown below, the  conceptual
difficulties which characterize the left eigenvalue problem
readily disappear by resorting to right eigenvalues. The need to
apply similarity transformations on $\mathbb{H}$, $\mathbb{C}$ and
$\mathbb{R}$ linear quaternionic matrices introduces {\em complex}
or {\em real} constraints on the right eigenvalues. The choice of
complex or real (right) eigenvalues will be extremely useful in
finding a practical method of resolution and manipulating
quaternionic matrices.

\subsection*{III-A. LEFT EIGENVALUES}

The left eigenvalue problem for $\mathbb{H}$ linear quaternionic
operators reads
\begin{equation}
\label{BB1}
\mathcal{O}_{\Ha} \psi = q \, \psi~,~~~\psi \in
V[n,\mathbb{H}]~,~q \in \mathbb{H}~.
\end{equation}
This problem  has been recently studied in the mathematical
literature~\cite{Hu00,HuSo01}. Nevertheless, no systematic way to
approach the problem has been given. We point out some difficulties
which appear in solving the left eigenvalue equation.\\

\noindent $\bullet$ {\em Similarity transformations}. In finding
the  solution  of Eq.~(\ref{BB1}), a first  difficulty is
represented by the impossibility to apply similarity
transformations, $S_{\Ha} \in M[n,\mathbb{H}]$, without losing the
formal structure of the left eigenvalue equation. In fact, by
observing that $S_{\Ha} \, q \neq q \, S_{\Ha}$, the quaternionic
matrices
\begin{equation}
M_{\Ha}~~~\mbox{and}~~~S_{\Ha} \,
M_{\Ha} \, S^{- 1}_{\Ha}
\end{equation}
do {\em not} necessarily satisfy the same eigenvalue equation.
Consequently, we can have quaternionic matrices with the same left
eigenvalue spectrum, but {\em no} similarity transformation
relating them. Explicit example are found in ref.~\cite{Del00}.\\

\noindent $\bullet$ {\em Hermitian operators}. Let
$\mathcal{O}_{\Ha}$ be an hermitian quaternionic operator,
$\psi$ the eigenvector corresponding to the eigenvalue $q$. By
using Eq.~(\ref{BB1}) and denoting by $\langle \varphi | \psi
\rangle$ the inner product in $V[n,\mathbb{H}]$, we obtain
\[
0 =  \langle \mathcal{O}_{\Ha} \psi | \psi \rangle - \langle \psi
| \mathcal{O}_{\Ha} \psi \rangle ~~~\Rightarrow~~~ 0 =
 \langle q \, \psi  | \psi \rangle - \langle \psi | q \,
\psi \rangle \neq (\bar{q} - q) \, \langle \psi| \psi \rangle~.
\]
Consequently,  the left eigenvalue problem for hermitian operators
could admit {\em quaternionic} solutions~\cite{M1}.\\

\noindent $\bullet$ {\em Square operators and eigenvalues}. As a
last difficulty in the use of left eigenvalues, we observe that if
$\psi $ is an $\mathcal{O}_{\Ha}$ eigenvector with eigenvalue $q$,
it will not necessarily be  an $\mathcal{O}_{\Ha}^{2}$ eigenvector
with eigenvalue $q^{2}$. In fact,
\[
 \mathcal{O}_{\Ha}^2 \psi = \mathcal{O}_{\Ha} \, q \, \psi
\neq q \, \mathcal{O}_{\Ha} \psi =
q^2 \, \psi~.
\]

\subsection*{III-B. RIGHT EIGENVALUES}

The right eigenvalue equation for $\mathbb{H}$-linear transformations reads
\begin{equation}
\label{B111}
\mathcal{O}_{\Ha}\psi =\psi \, q~,~~~
\psi \in V[n,\mathbb{H}]~,~q \in \mathbb{H}~.
\end{equation}
Such an equation can be reduced to a right complex eigenvalue
equation  rephasing the quaternionic eigenvalues by unitary
quaternions $u$,
\begin{equation}
\label{B8} \mathcal{O}_{\Ha} \psi \, u =\psi u \, \bar{u}qu=\psi u
\, z~,\qquad z\in \mathbb{C}~.
\end{equation}
This trick obviously fails for complex and real linear
transformations. In fact, due to the presence of the operators
$R_{i}$ in $\mathcal{O}_{\Co}$ and $\boldsymbol{R} \equiv \left(
R_{i},R_{j},R_{k} \right)$ in $\mathcal{O}_{\Re}$, we cannot apply
unitary transformations from the right. Observe that
\begin{equation}
\label{B10} (\mathcal{O}_{\Re,\Co}\psi ) \, u \neq
\mathcal{O}_{\Re,\Co}(\psi u)~, \qquad u\in \mathbb{H}~.
\end{equation}
The failure of the {\em associativity} in Eq.~(\ref{B10}) suggests
that we should consider {\em complex} eigenvalue equations for
$\mathbb{C}$-linear quaternionic operators,
\begin{equation}
\label{B11} \mathcal{O}_{\Co}\psi =\psi \, z~,\qquad z\in
\mathbb{C}~,
\end{equation}
and {\em real} eigenvalues for  $\mathbb{R}$-linear
quaternionic operators,
\begin{equation}
\label{B12}
\mathcal{O}_{\Re}\psi =\psi \, r~,\qquad r\in \mathbb{R}~.
\end{equation}
These equations are formally invariant under $\mathbb{C}$ and
$\mathbb{R}$-linear similarity transformations. Moreover, it can
easily be proved that
\begin{eqnarray*}
\mathcal{O}_{\Ha}^{n}\psi = \psi \, q^{n}~,~~~
\mathcal{O}_{\Co}^{n}\psi =\psi \, z^{n}~,~~~
\mathcal{O}_{\Re}^{n}\psi =\psi r^{n}~.
\end{eqnarray*}
It is important to note here that $\mathbb{R}$-linear quaternionic
operators admit real eigenvalues only in particular cases. Thus,
Eq.(\ref{B12}) has to be generalized. As shown later, a
satisfactory discussion of the eigenvalue problem for
$\mathbb{R}$-linear quaternionic operators will require the use of
a system of {\em coupled} equations.

\section*{IV. CANONICAL FORMS}

In this section, following the procedure introduced in the paper
of ref.~\cite{Del00}, we  discuss the canonical forms for
$\mathbb{H}$, $\mathbb{C}$ and $\mathbb{R}$-linear quaternionic
matrices. The results we will establish find an immediate
application in the theory of quaternionic differential operators.
In fact, by using the canonical form $J_{\X}$ of a given matrix
$M_{\X}$ we can readily obtain the exponential
\[ \exp \left[ M_{\X} \, x \right] = S_{\X}
\exp \left[ J_{\X} \, x \right] S_{\X}^{-1}\]
 and consequently,
avoiding tedious calculations, to solve quaternionic differential
equations with constant coefficients~\cite{DeDu}.

\subsection*{IV-A. $\mathbb{H}$-LINEAR MATRICES}

While matrices over commutative rings have gained much attention,
the literature on matrices with quaternionic entries is often
fragmentary. The main difficulty is that, due to the
noncommutative nature of quaternions, the standard method of
resolution breaks down. Consequently, finding eigenvalues and
canonical forms represents a more delicate problem. The recent
renewed interest in quaternionic matrix theory~\cite{Del00,Zha97}
and its applications~\cite{DeDu} shed new light on this intriguing
research field. To facilitate access to the individual topics, we
recall the main properties of  $\mathbb{H}$-linear quaternionic
matrices~\cite{REF1,REF2,REF3,REF4}
 and repeat the relevant theorems
from~\cite{Del00,Zha97,Lee47,Wie54} without proofs, thus making
our exposition self-contained.

In approaching the problem of diagonalization we have to consider
a right quaternionic eigenvalue equation. In fact, from
\begin{equation}
\label{ep1}
M_{\Ha} \psi_{k} = \psi_{k} q_{k}~,~~~k=1,...,n~,~~~\psi_{k}\in
V[n,\mathbb{H}]~,
\end{equation}
in the case $M_{\Ha}$ is diagonalizable, we immediately get the
following matrix equation
\[
M_{\Ha} S_{\Ha}[\psi_1,\psi_2,...,\psi_n] =
S_{\Ha}[\psi_1,\psi_2,...,\psi_n] D_{\Ha}~,
\]
where
\[ D_{\Ha}=\mbox{diag}[q_1,q_2,...,q_n]~
\]
and $S_{\Ha} = S_{\Ha}[\psi_1,\psi_2,...,\psi_n]$ is
defined by $Col_{k}(S_{\Ha})=\psi_k$.

Consequently, the diagonalization of the matrix $M_{\Ha}$
\[
M_{\Ha} = S_{\Ha} D_{\Ha} S_{\Ha}^{-1}
\]
is obtained by solving the corresponding right eigenvalue problem.
It is important to note here that we have {\em infinite} ways to
diagonalize a quaternionic matrix $M_{\Ha}$,
\[
\mbox{diag}[u_1,u_2,...,u_n] \,
\mbox{diag}[q_1,q_2,...,q_n] \,
\mbox{diag}[\bar{u}_1,\bar{u}_2,...,\bar{u}_n]~.
\]
Geometrically speaking this means that $\mbox{Im}[q]$ can
arbitrarily be fixed on the sphere of ray $|\mbox{Im}[q]|$. By a
particular choice of the unitary matrix
\[ U_{\Ha} = \mbox{diag}[u_1,u_2,...,u_n] \]
we can set a preferred space direction, for example the positive
$i$ axis, and consequently a  {\em complex} (positive) eigenvalue
spectrum.

Let us now briefly recall some properties of the eigenvalue
spectrum of $\mathbb{H}$ linear quaternionic matrices. By using
the symplectic decomposition of the matrix $M_{\Ha}$
\begin{equation*}
M_{\Ha}=M_{1}+jM_{2}~,~~~M_{1,2} \in M[n,\mathbb{C}]~,
\end{equation*}
 and the symplectic decomposition of the  vector $\psi$,
\begin{equation*}
\psi =\psi _{1}+j\psi _{2}~,~~ \psi _{1,2} \in V[n,\mathbb{C}]~,
\end{equation*}
we can rewrite Eq.(\ref{B8}) in the following (complex) form
\begin{equation}
\tilde{M}_{\Ha} \tilde{\psi} =z \, \tilde{\psi}~,
\end{equation}
where
\begin{equation}
\label{C16}
\tilde{M}_{\Ha}=\left(
\begin{array}{rr}
M_{1} & -M_{2}^{\ast } \\
M_{2} & M_{1}^{\ast }
\end{array}
\right) \in M[2n,\mathbb{C}]~~~\mbox{and}~~~
\tilde{\psi} =\left(
\begin{array}{c}
\psi _{1} \\
\psi _{2}
\end{array}
\right)\in V[2n,\mathbb{C}] .
\end{equation}
The mapping
\begin{equation}
\label{C18'}
f:M_{\Ha}\longmapsto \tilde{M}_{\Ha}
\end{equation}
is an isomorphism of the ring of quaternionic matrices $M_{\Ha}$
into the ring of the corresponding complex counterparts
$\tilde{M}_{\Ha}$. It is important to observe that this
isomorphism do not preserve the inner product of
eigenvectors~\cite{M1}. Nevertheless, the choice of a {\em complex
projection} of quaternionic inner products~\cite{ROT89} opens the
door to interesting applications in relativistic quantum
mechanics~\cite{DELFPL}. The {\em complex} orthogonality of
quaternionic eigenvectors (instead of a quaternionic
orthogonality) implies a doubling of solution in the
two-dimensional quaternionic Dirac equation~\cite{ROT89}. The {\em
four} (complex) orthogonal quaternionic solutions describe
particle/antiparticle with spin up/down. The use of {\em complex}
inner products is also a fundamental ingredient in the formulation
of gauge theories by geometric algebras~\cite{QF}.

The next theorem states the main property of the eigenvalue
spectrum of $\mathbb{H}$ linear quaternionic matrices,
for a detailed discussion see refs.~\cite{REF4,Lee47,Wie54}.\\

\noindent
\textbf{Theorem 1.}\\
\textit{Let }$\tilde{M}_{\Ha}$ \textit{be the matrix given in Eq.(\ref{C16}).
Then, its eigenvalues
appear in complex conjugate pairs.}\\

\noindent By using the result of Theorem 1 and the Gram-Schmidt
method, we can readily  obtain the
triangular form for $\mathbb{H}$-linear quaternionic matrices.\\

\noindent
\textbf{Theorem 2.}\\
\textit{Every }$M_{\Ha}$ \textit{is unitarily similar to an upper triangular
matrix.}\\

\noindent Moreover, a Jordan form can be given for
every $\mathbb{H}$-linear quaternionic matrix.\\

\noindent
\textbf{Theorem 3.}\\
\textit{Every }$n\times n$\textit{\ matrix with real quaternion elements is
similar under a matrix transformation with real quaternion elements to a
matrix in (complex) Jordan normal form with diagonal elements in the
complex field.}\\

\noindent To prove Theorem 3, we can use the isomorphism defined
in (\ref{C18'}). To any $\tilde{M}_{\Ha}$ corresponds a $2n\times
2n$ matrix $B$ in the (complex) Jordan form $B=B_{\Ha}\oplus
B_{\Ha}^{\ast }$, such that $\tilde{M}_{\Ha}P=PB$, where the
(non-singular) matrix $P$ has the form
\begin{equation*}
P=\left(
\begin{array}{rr}
P_{1} & -P_{2}^{\ast } \\
P_{2} & P_{1}^{\ast }
\end{array}
\right) .
\end{equation*}
This implies that
\begin{equation}
(P_{1}+jP_{2})^{-1}M_{\Ha}(P_{1}+jP_{2})=B_{\Ha}=D_{\Ha}+N_{\Ha}~,
\end{equation}
where $D_{\Ha}$ and $N_{\Ha}$ respectively denote the diagonal and
the nilpotent parts of $B_{\Ha}$.

\subsection*{IV-B. $\mathbb{C}$-LINEAR MATRICES}

Let us now consider
$\mathbb{C}$-linear transformations.
We can associate to any $n \times n$
$\mathbb{C}$-linear quaternionic matrix
a $2n$-dimensional complex matrix by the following mapping
\begin{equation}
\label{DD1}
M_{\Co}=M_{\Ha}+M_{\Ha}' R_{i}\longleftrightarrow
\tilde{M}_{\Co}=f(M_{\Ha})+i \, f(M_{\Ha}')=
\left(
\begin{array}{rr}
M_{1} & -M_{2}^{\ast } \\
M_{2} & M_{1}^{\ast }
\end{array}
\right)+
i \left(
\begin{array}{rr}
M_{1}' & -M_{2}^{'\ast } \\
M_{2}' & M_{1}^{'\ast }
\end{array}
\right) ,
\end{equation}
where $f$ denotes the isomorphism defined in (\ref{C18'}).
Then the following
proposition holds:\\

\noindent
\textbf{Proposition 1.}\\
\textit{Let $M_{\Co}$ be a $\mathbb{C}$-linear quaternionic
matrix and $\tilde{M}_{\Co}$ its complex counterpart
(see Eq. (\ref{DD1})). The mapping }
\begin{equation}
\label{DD11}
g:M_{\Co}\longmapsto \tilde{M}_{\Co}
\end{equation}
\textit{is an isomorphism of the ring of the $n$-dimensional
$\mathbb{C}$-linear matrices into the ring of $2n$-dimensional
complex matrices.}\\
Indeed, if
\begin{equation*}
A_{\Co}=A_{0,\Ha}+A_{1,\Ha}R_{i}\text{ \ and \ }
B_{\Co}=B_{0,\Ha}+B_{1,\Ha}R_{i}
\end{equation*}
are two $\mathbb{C}$-linear matrices, their corresponding complex
counterparts are given by
\begin{equation*}
g(A_{\Co})=f(A_{0,\Ha})+if(A_{1,\Ha}),\qquad
g(B_{\Co})=f(B_{0,\Ha})+if(B_{1,\Ha})
\end{equation*}
Then,
\begin{eqnarray*}
A_{\Co}B_{\Co}&=&C_{\Co}\\
&=&
A_{0,\Ha}B_{0,\Ha}+A_{1,\Ha}R_{i}B_{0,\Ha}+A_{0,\Ha}B_{1,\Ha}R_{i}+A_{1,\Ha}
R_{i}B_{1,\Ha}R_{i}\\
&=&
A_{0,\Ha}B_{0,\Ha}-A_{1,\Ha}B_{1,\Ha}+(A_{0,\Ha}B_{1,\Ha}+A_{1,\Ha}
B_{0,\Ha})R_{i}
\end{eqnarray*}
and
\begin{eqnarray*}
g(A_{\Co})g(B_{\Co}) & = &
f(A_{0,\Ha})f(B_{0,\Ha})-f(A_{1,\Ha})f(B_{1,\Ha})+
if(A_{1,\Ha})f(B_{0,\Ha})+i f(A_{0,\Ha})f(B_{1,\Ha})\\
&= &
f(A_{0,\Ha}B_{0,\Ha}-A_{1,\Ha}B_{1,\Ha})+
if(A_{0,\Ha}B_{1,\Ha}+A_{1,\Ha}B_{0,\Ha})\\
& =& g(C_{\Co}) \mbox{\tiny $\Box$}
\end{eqnarray*}

\noindent By using this isomorphism, the right eigenvalue spectrum
of  $\mathbb{C}$-linear quaternionic matrices can easily be
determined \cite{Del00}. The following result
\begin{center}
{\em ``A $\mathbb{C}$-linear matrix is diagonalizable
if the corresponding complex counterpart is diagonalizable''.}
\end{center}
was proven in Ref.~\cite{Del00} (where a preliminary  discussion
of the eigenvalue problem for $\mathbb{C}$-linear quaternionic
matrix operators was given). It is worth pointing out that the
converse of the previous statement is, in general, not true. For
instance, let us consider the complex matrix
\begin{equation}
\tilde{G}_{\Co}=\left(
\begin{array}{cccc}
z_{1} & 0 & 1 & 0 \\
0 & z_{2} & 0 & 1 \\
0 & 0 & z_{1} & 0 \\
0 & 0 & 0 & z_{2}
\end{array}
\right).
\end{equation}
This matrix admits a corresponding {\em diagonalizable}
$\mathbb{C}$-linear quaternionic matrix given by
\begin{equation}
G_{\Co}=D_{\Co}+N_{\Co}=\left(
\begin{array}{cc}
\mbox{Re}(z_{1})+\mbox{Im}(z_{1})R_{i} & 0 \\
0 & \mbox{Re}(z_{2})+\mbox{Im}(z_{2})R_{i}
\end{array}
\right) + \mbox{$\frac{1}{2}$} \,\left(
\begin{array}{cc}
-j+kR_{i} & 0 \\
0 & -j+kR_{i}
\end{array}
\right)~,
\end{equation}
where $N_{\Co}$ is nilpotent, diagonal and commutes with $D_{\Co}.$

The normal form of a $\mathbb{C}$-linear quaternionic matrix can
easily be calculated. Indeed, given any $\mathbb{C}$-linear
transformation $M_{\Co}$  and its corresponding complex
counterpart $\tilde{M}_{\Co}$ [see Eq.(\ref{DD1})],
 from the known properties of the Jordan form of complex matrices, we can
immediately obtain
\[
\tilde{S}_{\Co}^{-1} \tilde{M}_{\Co} \tilde{S}_{\Co}=
\tilde{J}_{\Co}= \tilde{D}_{\Co}+ \tilde{N}_{\Co}~,
\]
where $\tilde{D}_{\Co}$ is diagonal, $ \tilde{N}_{\Co}$ is
nilpotent and $[\tilde{D}_{\Co}, \tilde{N}_{\Co}]=0$. Then, the
quaternionic $\mathbb{C}$-linear matrices $M_{\Co}, S_{\Co},
D_{\Co},$ and  $N_{\Co}$ are uniquely determined by the
isomorphism stated in Proposition 1.

\subsection*{IV-C. $\mathbb{R}$-LINEAR MATRICES}

In the $n$ dimensional quaternionic vector space
$V[n,\mathbb{H}]$, the $\mathbb{R}$-linear transformations are
represented by
\begin{equation}
M_{\Re} =  \sum_{\mu=0}^{3} M_{\mu,\Ha} \, R_{\mu}~,
\end{equation}
where $M_{\mu,\Ha}$ represent $\mathbb{H}$ linear quaternionic
matrices and $R_{\mu}$ are the right acting operators defined in
the second section. Any $M_{\Re}$ is then characterized by
$16n^{2}$ real parameters. We can translate  $\mathbb{R}$-linear
$n\times n$\ matrices into equivalent $4n\times 4n$ real matrices,
and vice-versa, by the following translation rules:
\begin{eqnarray}
\label{E31}
R_{i} &\longleftrightarrow &\mathbf{I}= \left(
\begin{array}{cccc}
0 & -\mathbf{1}_{n} & 0 & 0 \\
\mathbf{1}_{n} & 0 & 0 & 0 \\
0 & 0 & 0 & \mathbf{1}_{n} \\
0 & 0 & -\mathbf{1}_{n} & 0
\end{array}
\right) ~, \nonumber \\
R_{j} & \longleftrightarrow & \mathbf{J}=
\left(
\begin{array}{cccc}
0 & 0 & -\mathbf{1}_{n} & 0 \\
0 & 0 & 0 & -\mathbf{1}_{n} \\
\mathbf{1}_{n} & 0 & 0 & 0 \\
0 & \mathbf{1}_{n} & 0 & 0
\end{array}
\right)~, \\
R_{k} &\longleftrightarrow &\mathbf{K}= \left(
\begin{array}{cccc}
0 & 0 & 0 & -\mathbf{1}_{n} \\
0 & 0 & \mathbf{1}_{n} & 0 \\
0 & -\mathbf{1}_{n} & 0 & 0 \\
\mathbf{1}_{n} & 0 & 0 & 0
\end{array}
\right)~, \nonumber
\end{eqnarray}
and
\begin{equation}
\label{E311}
M_{\mu,\Ha}=M_{0}+iM_{1}+jM_{2}+kM_{3}\longleftrightarrow \hat{M}_{\mu,\Ha}=
\left(
\begin{array}{rrrr}
M_{0} & -M_{1} & -M_{2} & -M_{3} \\
M_{1} & M_{0} & -M_{3} & M_{2} \\
M_{2} & M_{3} & M_{0} & -M_{1} \\
M_{3} & -M_{2} & M_{1} & M_{0}
\end{array}
\right) ,
\end{equation}
where $M_{\mu,\Ha} \in M[n,\mathbb{H}]$, $M_{0,...,3} \in
M[n,\mathbb{R}]$, and $\hat{M}_{\mu,\Ha} \in M[4n,\mathbb{R}]$. It
is easy to verify that $\mathbf{I}, \mathbf{J}, \mathbf{K}$
commute with $\hat{M}_{\mu,\Ha}$
\[
\left[ \mathbf{I}\, , \,\hat{M}_{\mu,\Ha} \right] = \left[
\mathbf{J}\, , \,\hat{M}_{\mu,\Ha} \right] =\left[ \mathbf{K}\, ,
\,\hat{M}_{\mu,\Ha} \right]=0
\]
 and
\[
\mathbf{I}^{2}= \mathbf{J}^{2}= \mathbf{K}^{2}=
\mathbf{K} \mathbf{J} \mathbf{I} = - \mathbf{1}.
\]
The following proposition holds.\\

\noindent
\textbf{Proposition 2.}\\
\textit{Let $M_{\Re}$ be a $\mathbb{R}$-linear matrix and
$\hat{M}_{\Re}$ its real counterpart; then the mapping}
\begin{equation*}
h:M_{\Re}=M_{0,\Ha}+M_{1,\Ha}R_{i}+M_{2,\Ha}R_{j}+M_{3,\Ha}R_{k} \longmapsto
\hat{M}_{\Re}=\hat{M}_{0,\Ha} + \hat{M}_{1,\Ha} \mathbf{I} +
\hat{M}_{2,\Ha} \mathbf{J} + \hat{M}_{3,\Ha} \mathbf{K}
\end{equation*}
\textit{is an isomorphism of the ring of the $n$-dimensional
$\mathbb{R}$-linear matrices
into the ring of $4n$-dimensional real matrices $ \hat{M}_{\Re}$.}\\
Observe that $\widehat{M_{\Ha}M_{\Ha}'}= \hat{M}_{\Ha}
\hat{M'}_{\Ha}$. Let
\[ A_{\Re}=A_{0,\Ha}+A_{1,\Ha}R_{i}+A_{2,\Ha}R_{j}+A_{3,\Ha}R_{k}~~~
\mbox{and}~~~
B_{\Re}=B_{0,\Ha}+B_{1,\Ha}R_{i}+B_{2,\Ha}R_{j}+B_{3,\Ha}R_{k}
\]
be two $\mathbb{R}$-linear quaternionic matrices. Their
corresponding real counterparts are given by
\begin{equation*}
h(A_{\Re})= \hat{A}_{0,\Ha}+\mathbf{I} \hat{A}_{1,\Ha}+
\mathbf{J} \hat{A}_{2,\Ha}+ \mathbf{K} \hat{A}_{3,\Ha}~~~
\mbox{and}~~~
h(B_{\Re})= \hat{B}_{0,\Ha}+\mathbf{I} \hat{B}_{1,\Ha}+
\mathbf{J} \hat{B}_{2,\Ha}+ \mathbf{K} \hat{B}_{3,\Ha}
\end{equation*}
Then,
\begin{eqnarray*}
A_{\Re}B_{\Re}
&=&C_{\Re}\\
&=&(A_{0,\Ha}B_{0,\Ha}-A_{1,\Ha}B_{1,\Ha}-A_{2,\Ha}B_{2,\Ha}
-A_{3,\Ha}B_{3,\Ha})+ \\
&&(A_{0,\Ha}B_{1,\Ha}+A_{1,\Ha}B_{0,\Ha}+A_{2,\Ha}B_{3,\Ha}-
A_{3,\Ha}B_{2,\Ha})R_{i}+ \\
&&(A_{0,\Ha}B_{2,\Ha}+A_{2,\Ha}B_{0,\Ha}+
A_{3,\Ha}B_{1,\Ha}-A_{1,\Ha}B_{3,\Ha})R_{j}+ \\
&&(A_{0,\Ha}B_{3,\Ha}+A_{3,\Ha}B_{0,\Ha}+
A_{1,\Ha}B_{2,\Ha}-A_{2,\Ha}B_{1,\Ha})R_{k}
\end{eqnarray*}
and
\begin{eqnarray*}
h(A_{\Re})h(B_{\Re}) &=&
(\hat{A}_{0,\Ha} \hat{B}_{0,\Ha}- \hat{A}_{1,\Ha}
\hat{B}_{1,\Ha}- \hat{A}_{2,\Ha} \hat{B}_{2,\Ha}- \hat{A}_{3,\Ha}
\hat{B}_{3,\Ha})+ \\
&&( \hat{A}_{0,\Ha} \hat{B}_{1,\Ha}+ \hat{A}_{1,\Ha}
\hat{B}_{0,\Ha}+ \hat{A}_{2,\Ha} \hat{B}_{3,\Ha}- \hat{A}_{3,\Ha}
\hat{B}_{2,\Ha})\mathbf{I}+ \\
&& ( \hat{A}_{0,\Ha} \hat{B}_{2,\Ha}+ \hat{A}_{2,\Ha}
\hat{B}_{0,\Ha}+ \hat{A}_{3,\Ha} \hat{B}_{1,\Ha}- \hat{A}_{1,\Ha}
\hat{B}_{3,\Ha})\mathbf{J}+ \\
&&(\hat{A}_{0,\Ha} \hat{B}_{3,\Ha}+ \hat{A}_{3,\Ha}
\hat{B}_{0,\Ha}+ \hat{A}_{1,\Ha} \hat{B}_{2,\Ha}- \hat{A}_{2,\Ha}
\hat{B}_{1,\Ha})\mathbf{K}\\
&=&h(C_{\Re}) \mbox{\tiny $\Box$}
\end{eqnarray*}

\noindent We now discuss the canonical forms of
$\mathbb{R}$-linear matrices. Let $A_{\Re}$ be an
$\mathbb{R}$-linear transformation,  $ \hat{A}_{\Re}$ its real
counterpart, $\left\{ \lambda _{1}+i\mu _{1},\lambda _{2}+i\mu
_{2},..., \lambda _{s}+i\mu _{s} \right\}$ the complex eigenvalues
of $ \hat{A}_{\Re}$, and $\left\{  \lambda_{2s+1},...,
\lambda_{4n} \right\}$ the real eigenvalues of $ \hat{A}_{\Re}$.
As well known \cite{Wil65,Bar}, there exists a real orthogonal
matrix $O$ such that
\[ \hat{J}_{\Re}=O \hat{A}_{\Re} O^{T}~,\]
where
\begin{eqnarray}
\label{xxx} \hat{J}_{\Re} & = & \left(
\begin{array}{ccccccc}
X_{1} &  &  &  &  &  &  \\
& X_{2} &  &  &  & P &  \\
&  & \ddots &  &  &  &  \\
&  &  & X_{s} &  &  &  \\
& 0 &  &  & \lambda _{2s+1} &  &  \\
&  &  &  &  & \ddots &  \\
&  &  &  &  &  & \lambda _{4n}
\end{array}
\right) \nonumber \\
& = & \left(
\begin{array}{ccccccc}
X_{1} &  &  &  &  &  &  \\
& X_{2} &  &  &  & 0 &  \\
&  & \ddots &  &  &  &  \\
&  &  & X_{s} &  &  &  \\
& 0 &  &  & \lambda _{2s+1} &  &  \\
&  &  &  &  & \ddots &  \\
&  &  &  &  &  & \lambda _{4n}
\end{array}
\right) +\left(
\begin{array}{ccccccc}
0 &  &  &  &  &  &  \\
& 0 &  &  &  & P &  \\
&  & \ddots &  &  &  &  \\
&  &  & 0 &  &  &  \\
& 0 &  &  & 0 &  &  \\
&  &  &  &  & \ddots &  \\
&  &  &  &  &  & 0
\end{array}
\right)\\
& = & \hat{D}_{\Re}+ \hat{N}_{\Re}~. \nonumber
\end{eqnarray}
In the previous equation, $X_{r}$ represents a $2\times 2$ real
matrix with eigenvalues $\lambda _{r}\pm i\mu _{r}$. An
appropriate choice of $O$ guarantees that
\begin{equation}
\label{rm}
X_{r}=\left(
\begin{array}{cc}
\lambda _{r} & -\mu _{r} \\
\mu _{r} & \lambda _{r}
\end{array}
\right) .
\end{equation}
Let us come back to the $\mathbb{R}$-linear transformation
$A_{\Re}$. By using the translation rules given in
Eqs.~(\ref{E31},\ref{E311}), we can immediately give its canonical
form
\[ J_{\Re}=D_{\Re}+N_{\Re}~.\]
In particular, the diagonal elements of $D_{\mathbb{R}}$
corresponding to the quaternionic translation of the  real blocks
\begin{eqnarray}
\hat{D}_{1,\Re} & =  &\left(
\begin{array}{cccc}
\lambda _{m} & -\mu _{m} & 0 & 0 \\
\mu _{m} & \lambda _{m} & 0 & 0 \\
0 & 0 & \lambda _{m+1} & -\mu _{m+1} \\
0 & 0 & \mu _{m+1} & \lambda _{m+1}
\end{array}
\right)~, \nonumber \\
 \hat{D}_{2,\Re} & = & \left(
\begin{array}{cccc}
\lambda _{m} & -\mu _{m} & 0 & 0 \\
\mu _{m} & \lambda _{m} & 0 & 0 \\
0 & 0 & \lambda _{m+1} & 0 \\
0 & 0 & 0 & \lambda _{m+2}
\end{array}
\right)~,\\
 \hat{D}_{3,\Re} & = & \left(
\begin{array}{cccc}
\lambda _{m} & 0 & 0 & 0 \\
0 & \lambda _{m+1} & 0 & 0 \\
0 & 0 & \lambda _{m+2} & 0 \\
0 & 0 & 0 & \lambda _{m+3}
\end{array}
\right)~, \nonumber
\end{eqnarray}
are respectively given by
\begin{eqnarray}
D_{1,\Re} & = &  \mbox{$\frac{1}{2}$} \, [\lambda
_{m}(1-L_{i}R_{i})+\mu
_{m}(L_{i}+R_{i})+ \nonumber \\
 & & ~~~\lambda _{m+1}(1+L_{i}R_{i})+\mu _{m+1}(L_{i}-R_{i})]~,
\nonumber \\
D_{2,\Re} & = & \mbox{$\frac{1}{2}$} \, [\lambda
_{m}(1-L_{i}R_{i})+\mu
_{m}(L_{i}+R_{i})+ \nonumber \\
 & & ~~~\mbox{$\frac{1}{2}$} \, \lambda
_{m+1}(1+L_{i}R_{i}-L_{j}R_{j}+L_{k}R_{k})+ \mbox{$\frac{1}{2}$}
\,  \lambda _{m+2}(1+L_{i}R_{i}+L_{j}R_{j}-L_{k}R_{k})]~,\\
D_{3,\Re} & = & \mbox{$\frac{1}{4}$} \, [\lambda
_{m}(1-L_{i}R_{i}-L_{j}R_{j}-L_{k}R_{k})+\lambda
_{m+1}(1-L_{i}R_{i}+L_{j}R_{j}+L_{k}R_{k})+ \nonumber \\
 & & ~~~\lambda _{m+2}(1+L_{i}R_{i}-L_{j}R_{j}+L_{k}R_{k})+\lambda
_{m+3}(1+L_{i}R_{i}+L_{j}R_{j}-L_{k}R_{k})]~. \nonumber
\end{eqnarray}
\noindent
 As happens for $\mathbb{C}$-linear quaternionic matrices, an
$\mathbb{R}$-linear quaternionic matrix is diagonalizable if the
corresponding real counterpart is diagonalizable. The converse is
 not necessarily true.

\section*{V. THE EIGENVALUE PROBLEM FOR $\mathbb{R}$-LINEAR MATRICES}

Let us now consider the eigenvalue problem for $\mathbb{R}$-linear
quaternionic matrices. Eq.\,(\ref{B12}) is obviously too
restrictive. In fact, such an equation sets the real eigenvalue
spectrum of $\mathbb{R}$-linear quaternionic operators. No
information is given about the remaining eigenvalues. In
particular, if the real counterpart $\hat{M}_{\Re}$ of the
$\mathbb{R}$-linear quaternionic  matrix $M_{\Re}$ does not have
real eigenvalues, Eq.\,(\ref{B12}) does not admit solution. This
is very embarrassing if we consider, for example,
$\mathbb{R}$-linear anti-hermitian quaternionic operators. Thus,
we need to modify Eq.\,(\ref{B12}). The discussion regarding the
``pseudo-triangular'' form of the matrices $\hat{A}_{\Re}$ [see
Eq. (\ref{xxx})] suggests as $\mathbb{R}$-linear eigenvalue
problem the following  system of {\em coupled} equations
\begin{equation}
\label{E35}
\begin{array} {lcl}
M_{\Re}\psi & = & a \, \psi + b \, \varphi \\
 M_{\Re}\varphi & = & c \, \varphi +d \, \psi  \end{array}
\end{equation}
where
\[
a,b,c,d\in \mathbb{R}~, ~~
\psi=\psi_{0}+i\psi_{1}+j\psi_{2}+k\psi_{3}~,~~
\varphi=\varphi_{0}+i\varphi_{1}+j\varphi_{2}+k\varphi_{3}~,~~
\psi_{0,...,3},\varphi_{0,...,3} \in V[n,\mathbb{R}]~.
\]
It can be shown that the real coefficients $a,b,c,d$ are related
to  the real and imaginary part of the $\hat{M}_{\Re}$
eigenvalues. In fact, by translating the system  (\ref{E35}) into
its real matrix counterpart, we find
\begin{equation}
\label{E36}
\left(
\begin{array}{cc}
\hat{M}_{\Re} & 0 \\
0 & \hat{M}_{\Re}
\end{array}
\right) \left(
\begin{array}{c}
\hat{\psi} \\
\hat{\varphi}
\end{array}
\right) =\left(
\begin{array}{cc}
a \boldsymbol{1}_{4n} & b \boldsymbol{1}_{4n} \\
c \boldsymbol{1}_{4n} & d \boldsymbol{1}_{4n}
\end{array}
\right) \left(
\begin{array}{c}
\hat{\psi} \\
\hat{\varphi}
\end{array}
\right),
\end{equation}
where
\begin{equation*}
\hat{\psi}= \left(
\begin{array}{c}
\psi _{0} \\
\psi _{1} \\
\psi _{2} \\
\psi _{3}
\end{array}
\right)~,~~
 \hat{\varphi}= \left(
\begin{array}{c}
\varphi _{0} \\
\varphi _{1} \\
\varphi _{2} \\
\varphi _{3}
\end{array}
\right)~~\in V[4n,\mathbb{R}]~.
\end{equation*}
The matrix equation (\ref{E36}) admits non trivial solutions if
and only if
\begin{equation}
\label{E37}
\det \left[ \left(
\begin{array}{cc}
\hat{M}_{\Re} -a {\bf 1}_{4n} & -b {\bf 1}_{4n} \\
-c {\bf 1}_{4n} & \hat{M}_{\Re} -d {\bf 1}_{4n}
\end{array}
\right) \right] =0~.
\end{equation}
By rewriting the  matrix $\hat{M}_{\Re}$ in terms of the
symilarity matrix $\hat{S}_{\Re}$ and of its Jordan form
$\hat{J}_{\Re}$, i.e.
\[
\hat{M}_{\Re} =  \hat{S}_{\Re} \hat{J}_{\Re} \hat{S}^{-1}_{\Re}~,
\]
and by using  the cyclic property of the determinant, we reduce
Eq.\.(\ref{E37}) to
\begin{equation}
\det \left[ \left(
\begin{array}{cc}
\hat{J}_{\Re}-a {\bf 1}_{4n} & -b {\bf 1}_{4n} \\
-c {\bf 1}_{4n} & \hat{J}_{\Re}-d {\bf 1}_{4n}
\end{array}
\right) \right]=0~.
\end{equation}
By simple algebraic manipulations~\cite{Bar}, we obtain
\begin{eqnarray*}
\det \left[ \left(
\begin{array}{cc}
\hat{J}_{\Re}-a {\bf 1}_{4n} & -b {\bf 1}_{4n} \\
-c {\bf 1}_{4n} & \hat{J}_{\Re}-d {\bf 1}_{4n}
\end{array}
\right) \right] & = &  \det [(\hat{J}_{\Re}-a {\bf
1}_{4n})(\hat{J}_{\Re}-d {\bf 1}_{4n})-bc {\bf 1}_{4n}] \\
& = & \det [\hat{J}_{\Re}^{2}-(a+d) \hat{J}_{\Re}+(ad-bc) {\bf
1}_{4n}] \\ & =  & \prod_{i}[z_{i}^{2}-(a+d)z_{i}+(ad-bc)] \\
 & = & 0~,
\end{eqnarray*}
where $z_{i}$ represent the eigenvalues of the real matrix
$\hat{M}_{\Re}$. The previous equation explicitly shows the
relation between the real coefficients $a,b,c,d$ (which appear in
the $\mathbb{R}$-linear eigenvalue problem) and the eigenvalues
$z$ of the real counterpart of the quaternionic matrix $M_{\Re}$.
In the case of complex eigenvalues $z$, we find
\begin{equation}
\label{E40}
\Delta =(a+d)-4(ad-bc)=(a-d)^{2}+bc<0~.
\end{equation}
This condition guarantees that the  eigenvalues of the real matrix
\begin{equation*}
Z=\left(
\begin{array}{cc}
a {\bf 1}_{4n} & b {\bf 1}_{4n} \\
c {\bf 1}_{4n} & d {\bf 1}_{4n}
\end{array}
\right)
\end{equation*}
appear in conjugate pairs. Consequently, we can find a real
similarity transformation $T$ such that
\begin{equation*}
T \,Z \, T^{-1}=\left(
\begin{array}{cc}
\lambda {\bf 1}_{4n} & -\mu {\bf 1}_{4n} \\
\mu {\bf 1}_{4n} & \lambda {\bf 1}_{4n}
\end{array}
\right)~.
\end{equation*}
Finally, without loss of generality, we can consider the following
eigenvalue problem for $\mathbb{R}$-linear transformations
\begin{eqnarray}
\label{R1}
\begin{array}{lcl}
M_{\Re}\psi & = &\lambda \, \psi -\mu \, \varphi~, \\
M_{\Re}\varphi &=&\lambda \, \varphi + \mu \, \psi ~.  \end{array}
\end{eqnarray}

\section*{VI. FINAL REMARKS}

These final remarks aim to give a concluding discussion on the
``coupled'' eigenvalue problem and a brief summary of mathematical
and physical applications  motivating our interest in this
research. In particular, we are interested to bring together two
areas: quaternionic differential operators and quantum mechanics.

\subsection*{Coupled eigenvalue equations}

In the previous Section, we have introduced, for
$\mathbb{R}$-linear transformations, the eigenvalue problem
(\ref{R1}) which represents the {\em natural} generalization of
(\ref{B12}). In particular, as we observed above, the study of
system (\ref{R1}) instead of Eq.(\ref{B12}) allows to take into
account the existence of complex eigenvalues and, consequently,
complete the eigenvalue spectrum of $\mathbb{R}$-linear
quaternionic operators. Actually, the eigenvalue problem
(\ref{R1}) also applies to $\mathbb{H}$ and $\mathbb{C}$-linear
transformations.  It can be considered as an equivalent
formulation of Eqs.(\ref{B111}) and (\ref{B11}). To show that, let
us  consider the equation
\begin{equation}
\label{RR1}
M_{\Co} \psi = \psi z = \psi \lambda + \psi i \mu~.
\end{equation}
We limit ourselves to discuss $\mathbb{C}$-linear transformations.
Obviously, if a preferred complex direction is chosen for the
eigenvalues of $\mathbb{H}$-linear quaternionic operators, all the
arguments in what follows  also hold for $\mathbb{H}$-linear
transformations. By using the $\mathbb{C}$-linearity, we find
\[
(M_{\Co} \psi)i=M_{\Co}(\psi i)=\psi i \lambda - \psi \mu~.
\]
The pair of eigenvector $(\psi,\varphi=\psi\, i)$, where $\psi$ is
solution of  Eq.(\ref{RR1}),  satisfies the coupled equations
\begin{eqnarray}
\label{R2}
\begin{array}{lcl}
M_{\Co}\psi &=&\lambda \, \psi -\mu \, \varphi~,\\
M_{\Co}\varphi & =&\mu \psi + \lambda \varphi~.  \end{array}
\end{eqnarray}
Vice-versa, let $M_{\Co}$ and (\ref{R2}) be respectively a
$\mathbb{C}$-linear transformation and the corresponding
eigenvalue problem. We denote by $(\psi,\varphi)$ a solution of the system
(\ref{R2}). If $\psi$ satisfies Eq.(\ref{RR1}) too, then,
comparing Eq.(\ref{RR1}) and the first equation in (\ref{R2}), one
immediately obtains $\varphi=\psi i$. If, on the contrary, $\psi$
is not a solution of Eq.(\ref{RR1}), by using the
$\mathbb{C}$-linearity, we obtain
\[
 M_{\Co} (\varphi i)=(M_{\Co} \varphi) i= \mu \psi i + \varphi i~.
\]
Thus,
\begin{equation}
M_{\Co}(\psi + \varphi i)= (\psi + \varphi i)z.
\end{equation}
Hence, it is possible to associate to  any solution
$(\psi,\varphi)$ of the system (\ref{R2}) a corresponding
eigenvector of $M_{\Co}$.

It is worth pointing out that the coupled system (\ref{R1}) can be
obtained by solving the eigenvalue problem (\ref{B12}) for   {\em
complexified} quaternionic eigenvectors and {\em complexified}
real eigenvalues. Infact, by imposing that
\[ \psi \to \Psi = \psi + I \, \varphi~~~\in~ \mathbb{H}(1,i,j,k) \, \otimes
\, \mathbb{C}(1,I)~,
\]
and
\[
r \to Z = \lambda + I \mu~~~\in~\mathbb{C}(1,I)~,
\]
from the {\em complexified} eigenvalue problem (\ref{B12}),
\[
M_{\Re} \, \Psi = \Psi \, Z~,
\]
 we immediately get the
coupled system (\ref{R1}).

\subsection*{Applications}

Many physical problems dealing with differential operators are
greatly simplified by using the matrix formalism and solving the
corresponding eigenvalue problem.

Let us first consider a very simple case. That is the
$\mathbb{H}$-linear second order homogeneous ordinary differential
equation
\begin{equation}
\label{hde} \ddot{\psi}(x) - \alpha \, \dot{\psi}(x) - \beta \,
\psi(x) = 0~,~~~~\alpha,\beta \in \mathbb{H}~,~x \in \mathbb{R}~.
\end{equation}
In looking for quaternionic exponential solution
$\psi(x)=\exp[q\,x]$ and observing that the derivative of
 $\exp[q\,x]$ with respect to the real variable $x$ is  $q \,\exp[q\,x]$,
  we reduce the previous problem
to find the solutions of the following quadratic equation
\begin{equation}
\label{pe2} q^{2} = \alpha \, q + \beta~.
\end{equation}
This equation can be rewritten in matrix form as follows
\begin{equation}
\label{mf} M_{\Ha} \left( \begin{array}{c} q \\ 1 \end{array}
\right) = \left( \begin{array}{c} q^2 \\ q \end{array}
\right)~,~~~M_{\Ha} = \left( \begin{array}{cc} \alpha & \beta\\ 1
& 0 \end{array} \right)~.
\end{equation}
As seen in this paper, the $\mathbb{H}$-linear quaternionic matrix
$M_{\Ha}$ satisfies a right (complex) eigenvalue equation
\begin{equation}
\label{mf2} M_{\Ha} \left( \begin{array}{c} v \\ w \end{array}
\right) = \left( \begin{array}{c} v \\ w \end{array} \right) \,
z~,~~~z \in \mathbb{C}~,~v,w \in \mathbb{H}~,~w\bar{w}=1~.
\end{equation}
Due to the particular form of $M_{\Ha}$, the components of the
$M_{\Ha}$-eigenvectors satisfy the following condition
\begin{equation}
\label{con}
v=w \, z~.
\end{equation}
Multiplying (from the right) Eq.\,(\ref{mf2}) by $\bar{w}$ and using
the constraint (\ref{con}), we obtain
\begin{equation}
\label{mf3}
M_{\Ha} \left( \begin{array}{c} w z \bar{w} \\ 1
\end{array} \right) = \left( \begin{array}{c} w z^2 \bar{w}  \\ w z \bar{w}
\end{array}
\right)~.
\end{equation}
Comparing Eq.\,(\ref{mf}) with Eq.\,(\ref{mf3}), we immediately
get
\begin{equation}
\label{sol}
q = w \, z \, \bar{w}~.
\end{equation}
The problem of finding exponential solutions for
$\mathbb{H}$-linear differential with constant coefficients and,
consequently, zeros of $\mathbb{H}$-linear polynomial
equations~\cite{SER}, is thus equivalent to solve the right
(complex) eigenvalue problem for the associated matrix. Obviously,
the previous considerations also hold for the $n$-dimensional
case.

The solutions of  $\mathbb{X}$-linear quaternionic differential
equations with constant coefficients
\begin{equation}
\label{xeq} \psi^{(n)}(x) - A_{n-1,\X} \psi^{(n-1)}(x) -
A_{n-2,\X} \psi^{(n-2)}(x) - ... - A_{0,\X} \psi(x) = 0~,~~~
\mathbb{X}=\mathbb{R},\mathbb{C},\mathbb{H}~.
\end{equation}
can be given in terms of the eigenvalues and eigenvectors of the
matrix
\[
\left( \begin{array}{cccccc} A_{n-1,\X} & A_{n-2,\X} & . & . & . &
A_{0,\X} \\ 1 & 0 & . & . & . & 0 \\
 \cdot &   \cdot  &  \cdot &  \cdot & \cdot & \cdot \\
 \cdot &   \cdot  &  \cdot &  \cdot & \cdot & \cdot \\
  \cdot &   \cdot  &  \cdot &  \cdot & \cdot & \cdot \\
 0 & 0 & . & . & 1 & 0
\end{array} \right)
\]
Interesting $\mathbb{C}$-linear differential equations appear in
quaternionic quantum mechanics~\cite{ADL}. For example, by
studying quaternionic tunneling effect as candidate to possible
phenomenological  deviations  from the standard (complex) theory,
we have to solve the following $\mathbb{H}$-linear Schr\"odinger
equation
\begin{equation}
\label{se1}
\partial_{t} \Psi(x,t) = \left[ \mbox{$\frac{i}{\hbar}$} \,
 \left( \mbox{$\frac{\hbar^{\2}}{\2m}$} \, \partial_{\x \x} - i \, V
\right) + \mbox{$\frac{j}{\hbar}$} \, W \right]  \Psi(x,t)
\end{equation}
where $\mbox{$\frac{j}{\hbar}$} \, W$ represents the {\em new}
quaternionic perturbation. The quaternionic stationary state wave
function
 \begin{equation}
\label{solw}
\Psi(x,t) = \psi(x) \, \exp \left[ - \mbox{$\frac{i}{\hbar}$} \, E t
\right]
\end{equation}
is solution of Eq.\,(\ref{se1}) on the condition that $\psi(x)$
be solution of the following time-independent  $\mathbb{C}$-linear
(ordinary) differential equation
\begin{equation}
\label{ose}
i \,  \mbox{$\frac{\hbar^{\2}}{\2m}$} \, \ddot{\psi}(x) - i \, V \,
\psi(x) +  j \, W \,  \psi(x) + \psi(x) \, i \, E = 0~.
\end{equation}
Observe that the choice of the imaginary unit $i$ in the Laplacian
operator $\partial_{\x \x}$, Eq.(\ref{se1}), and in the
 time exponential,   Eq.(\ref{solw}), is fundamental
 to recover the standard
results in the complex limit. In this formalism,  quaternionic
potentials are treated as perturbation effects on standard quantum
mechanics. We also point out that the right position of the time
exponential is fundamental to perform the separation of variables.

The solution of Eq.(\ref{ose}) can be given in terms of the eigenvalues and
eigenvectors of the $\mathbb{C}$-linear matrix
\[
\left( \begin{array}{cc} 0 & A_{0,\Co}\\ 1 & 0
\end{array} \right)~,
\]
where
\[ A_{0,\Co} = \mbox{$\frac{2m}{\hbar^{\2}}$} \, \left(
V + L_k \, W + L_i \, R_i \, E \right)~.
\]
A detailed phenomenological discussion of the quaternionic
tunneling effect is found in the paper of ref.~\cite{DeDuNi}.

\subsection*{Outlooks}

As seen in this paper, the choice of right (complex) eigenvalues
for $\mathbb{H}$ and $\mathbb{C}$ linear operators play a
fundamental role in  discussing canonical forms, in finding
solutions of polynomial and differential equations. It was shown
that the right (complex) eigenvalue problem is equivalent to a
``coupled'' system and this was extremely  important to study the
eigenvalue problem for  $\mathbb{R}$-linear quaternionic matrices,
where a pair of real eigenvalues must be introduced. This work was
intended as an attempt at motivating the study of $\mathbb{R}$ and
$\mathbb{C}$-linear quaternionic operators in view of possible
applications in quantum mechanics and gauge theory. It would be
desirable to give a complete theory of $\mathbb{X}$-linear
quaternionic matrices and differential operators. More
realistically, this paper touches only a few aspects of the theory
and shows how the choice of the right eigenvalue equation seems to
be the best to investigate quaternionic formulations of physical
theory. It was not our purpose to study here differential
operators. The results in this field are far from being conclusive
and some questions represent at present intriguing challenges:
 variations of parameters;
 order reduction;
not invertible higher derivative $\mathbb{R}$ and
$\mathbb{C}$-linear constant coefficients;  variable coefficients;
 integral
transforms. Finally, it would be desirable to extend the
discussion on the eigenvalue problem by matrix translation to the
non-associative case~\cite{KHA,M2,M3}

\subsection*{Acknowledgments}
The authors wish to express their thanks to Profs. N. Cohen and G.
Ducati for the helpful discussions during the preparation of the
paper and for drawing their attention to some interesting
applications in matrix and differential operator theory. The
authors also thank an anonymous referee for comments, references
and suggestions. One of the authors (SDL) gratefully acknowledges
the University of Lecce (Department of Physics) and Curitiba
(Department of Mathematics) for the hospitality, and the FAEP
(State University of Campinas) for financial support.

\end{document}